\documentclass{aa}
\usepackage{graphics,psfig}

\begin{document}

%
   \title{The line--of--sight distribution of the gas in the inner 60~pc of the Galaxy}

   \author{B.~Vollmer\inst{1}, R.~Zylka\inst{2,4} \and  W.J.~Duschl\inst{3,1}}

   \offprints{B.~Vollmer, e-mail: bvollmer@mpifr-bonn.mpg.de}

   \institute{Max-Planck-Institut f\"ur Radioastronomie, Auf dem H\"ugel 69,
   	      53121 Bonn, Germany. \and
	      Phys. Institut d. Universit\"{a}t zu K\"{o}ln, Z\"{u}lpicher Str. 77, 
	      50937 K\"{o}ln, Germany \and 
	      Institut f\"ur Theoretische
	      Astrophysik der Universit\"at Heidelberg, Tiergartenstra{\ss}e 15,
              69121 Heidelberg, Germany \and
	      IRAM, 300, rue de la piscine, 38406 Saint Martin d'H\`eres, France.}

   \date{Received / Accepted}

   \authorrunning{Vollmer et al.}
   \titlerunning{The LOS distribution in the inner 60~pc of the Galaxy}

\abstract{
2MASS K$_{\rm S}$ band data of the inner 60~pc of the Galaxy are used to reconstruct 
the line-of-sight distances of the giant molecular clouds located in this region.
Using the 2MASS H band image of the same region, 
two different populations of point sources are identified
according to their flux ratio in the two bands. The population of blue point 
sources forms a homogeneous foreground that has to be subtracted before analyzing
the K$_{\rm S}$ band image. The reconstruction is made using two basic assumptions:
(i) an axis-symmetric stellar distribution in the region of interest and (ii)
optically thick clouds with an area filling factor of $\sim$1 
that block all light of stars located behind them.
Due to the reconstruction method, the relative distance between the different cloud 
complexes is a robust result, whereas it is not excluded that the absolute distance with
respect to Sgr~A$^{*}$
of structures located more than 10~pc in front of Sgr~A$^{*}$ are understimated by up to a 
factor of 2. It is shown that all structures observed in the 1.2~mm continuum and in 
the CS(2-1) line are present in absorption. We place the 50~km\,s$^{-1}$ cloud complex close to, 
but in front of, Sgr~A$^{*}$. The 20~km\,s$^{-1}$ cloud complex is located in front of the
50~km\,s$^{-1}$ cloud complex and has a large LOS distance gradient along the direction of 
the galactic longitude. The bulk of the Circumnuclear Disk is not seen in absorption. 
This leads to an upper limit of the cloud sizes within the Circumnuclear Disk of $\sim$0.06~pc.
}

\maketitle

\section{Introduction}

The Galactic Center (GC) is a unique place to study the fueling of a central
black hole in great detail. The black hole, which coincides with the non-thermal 
radio continuum source Sgr~A$^{*}$, has a mass of 
$\sim$3\,10$^{6}$~M$_{\odot}$ (Eckart \& Genzel 1996). At a distance between
2~pc and 7~pc a ring-like structure made of distinct clumps, which is called
the Circumnuclear Disk (CND) is rotating around the central point mass.
The inner ionized edge of the CND is a part of a structure of ionized gas 
(see e.g. Lo \& Claussen 1983, Lacy et al. 1991) that resembles a spiral and 
is therefore called the Minispiral. Sgr~A$^{*}$ is surrounded by a huge 
H{\sc ii} region Sgr~A West with a size\footnote{We assume 8.5 kpc for 
the distance to the Galactic Centre} of 2.1$\times$2.9 pc, which was first 
observed by Ekers et al. (1975). 

The Circumnuclear disk has been observed in several molecular lines:
Gatley et al. (1986) (H$_{2}$), Serabyn et al. (1986) (CO,CS), G\"{u}sten
et al. (1987) (HCN), DePoy et al. (1989) (H$_{2}$), Sutton et al. (1990)
(CO), Jackson et al. (1993) (HCN), Marr et al. (1993) (HCN),
Coil \& Ho (1999, 2000) (NH$_{3}$), and Wright et al. (2001) (HCN).
The deduced properties of the CND are the following:
it has a mass of a few 10$^{4}$~M$_{\odot}$; 
the ring is very clumpy with an estimated volume filling factor of 
$\Phi_{\rm V} \sim 0.01$ and an area filling factor of $\Phi_{\rm A} \sim 0.1$;
the clumps have masses of $\sim$30~M$_{\odot}$, sizes of $\leq$0.1~pc, and
temperatures $\geq$100~K. 

These observations together with mm continuum observations 
(see, e.g. Mezger et al. 1989, Dent et al. 1993) have shown that
three giant molecular clouds (GMCs) are located in the inner 60~pc of the Galactic Center.
Following Zylka et al. (1990) these are: 
(i) Sgr~A East Core, a compact giant molecular cloud with
a gas mass of several 10$^{5}$ M$_{\odot}$ that is located to the north--east of
Sgr~A$^{*}$. (ii) The giant molecular cloud M-0.02-0.07 that is located to the
east of Sgr~A$^{*}$. Since its main radial velocity is $\sim$50~km\,s$^{-1}$,
it is also called the {\it 50~km\,s$^{-1}$ cloud}. (iii) The GMC complex M-0.13-0.08
 that is located to the south of Sgr~A$^{*}$. Since its main radial velocity 
is $\sim$20~km\,s$^{-1}$, it is also called the {\it 20~km\,s$^{-1}$ cloud}.
Coil \& Ho (1999, 2000), and Wright et al. (2001) argue that
there are physical connections between these GMC complexes and the CND on the grounds
of radial velocities, projected distances, and linewidths.
Their analysis misses crucial information, i.e. the gas distribution in the
line--of--sight (LOS). With this information it is possible to place the
GMC complexes in three--dimensional space and confirm or exclude
possible connections with the CND. 

In this article we use 2MASS near infrared data that show the
GMC complexes in absorption to reconstruct the LOS distribution of the gas
in the inner 60~pc of the Galactic Center. We use the absorption features in 
the NIR continuum emission whose level over a large scale ($\sim 13'$)
is very difficult to obtain.

This article has the following structure: the 2MASS data are presented in 
Sect.~\ref{sec:data} followed by the description of the data reduction
(Sect.~\ref{sec:data_reduc}). The final K$_{\rm S}$ band images are 
presented in Sect.~\ref{sec:results}. We explain the method of the reconstruction of
the line-of-sight distances and show its results in Sect.~\ref{sec:reconstruct}.
The reconstructed line-of-sight distribution is discussed and compared to 
mm observations in Sect.~\ref{sec:discussion}. We give our  
conclusions in Sect.~\ref{sec:conclusions}.

\section{The data \label{sec:data}}

We use the near--infrared (NIR) data of the {\it Two Micron All Sky Survey (2MASS)}. 
Six uncompressed full--resolution Atlas images in J (1.25~$\mu$m), H (1.65~$\mu$m), 
and K$_{\rm S}$ (2.17~$\mu$m) covering the inner
$30' \times 20'$ of the Galactic Center (Fig.~\ref{fig:mosaic}) were downloaded 
via the 2MASS Batch Image Server on the IRSA site.  
Single images have sizes of 512$\times$1024 pixels. Each pixel has a size of $1''$. The
images are overlapping over a region of 53 pixels in declination and
89 pixels in right ascension. The effective resolution of these
observations was $\sim 3''$.

\section{Data reduction \label{sec:data_reduc}}
 
\subsection{NIR continuum}

\begin{figure}
	\psfig{file=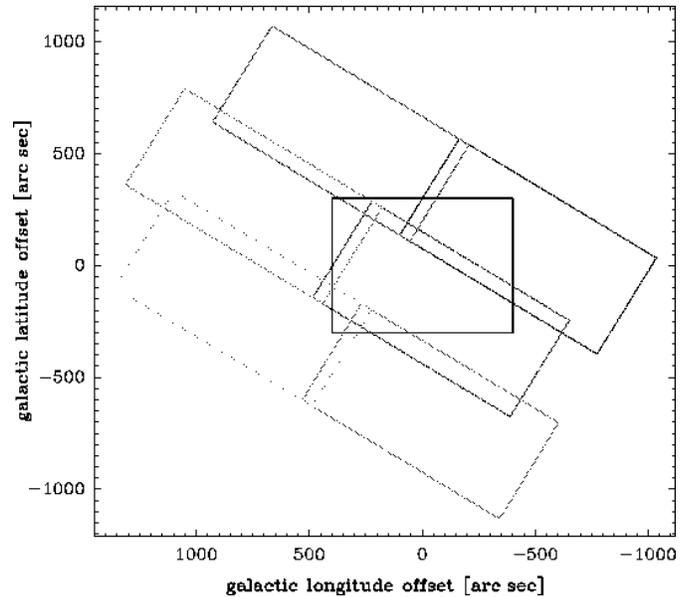,angle=-90,width=\hsize}
      \caption{Mosaic of 2MASS images centered on Sgr~A$^{*}$. 
	The thick box delineates our region of interest. 
	\label{fig:mosaic}} 
\end{figure}

All images were analyzed using the MOPSI\footnote{MOPSI is an astronomical
data reduction software developed by R. Zylka (see http://www.iram.fr/IRAMES/index.html).
} software. Since we are interested in the stellar
continuum, all 6 images had to be put together properly with special care 
taken for the sky and system emission correction. 
We used the inner image, which includes the Galactic Center, as the 
reference image. Fig.~\ref{fig:mosaic} shows the configuration of the
6 2MASS images in the sky.
In a first step the constant offset between the reference image
and a second, neighbouring image was calculated.
It turned out that these zero order corrections gave unsatisfactory results.  
Therefore, we used the overlapping region to correct the neighbouring 
image for a constant tilt.
All 5 surrounding images were treated in this way, in order to ensure smooth 
transitions between the images and thus flat baselines for the stellar continuum 
emission. At the end an image of $800'' \times 600''$ centered on
Sgr~A$^{*}$ was used.

\subsection{Single stars}

The images in all wavelengths contain foreground stars to different extents.
Following Launhardt et al. (2002) we divide the foreground star populations
into 4 classes: (i) Galactic Disk stars, (ii) Galactic Bulge stars,
(iii) Nuclear Stellar Disk, and (iv) Nuclear Stellar Cluster stars.
The Nuclear Stellar Disk and the Nuclear Stellar Cluster form the Nuclear
Bulge. The dereddend COBE fluxes at 2.2~$\mu$m (with a resolution of 
0.7$^{\rm o} \simeq$ 100~pc) of the Galactic and the Nuclear Bulge are of the same 
order, whereas the emission of the Galactic Disk is negligible (Launhardt et al. 2002). 
Philipp et al. (1999) estimated that more than 80\% of the integrated flux
density of the inner 30~pc is contributed by stars located in the Nuclear Bulge.

The overall populations of low and intermediate-mass main sequence stars
in the Nuclear and Galactic Bulge are similar, but the central 30~pc have an 
overabundance of K-luminous giants. These giants are more concentrated towards the
center than low-mass main sequence stars (Philipp et al. 1999). The near infrared
luminosity of the central 30~pc is dominated by these evolved stars,
whose contribution to the total stellar mass is however negligible (Mezger et al. 1999).

\begin{figure}
	\resizebox{\hsize}{!}{\includegraphics{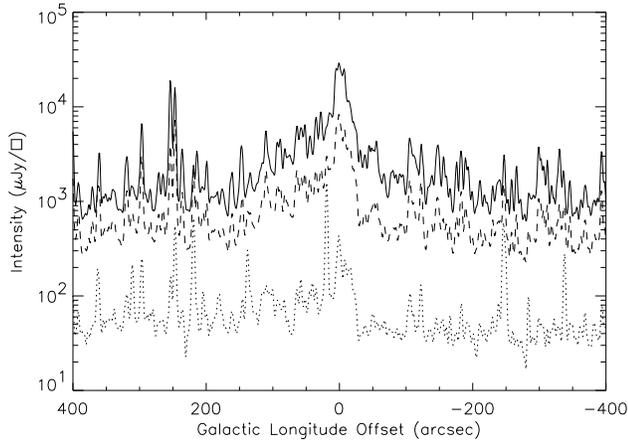}}
      \caption{Slice of the 2MASS images through Sgr~A$^{*}$ parallel to the galactic
	longitude. Solid line: K$_{\rm S}$ band. Dashed line: H band.
	Dotted line: J band. 
	\label{fig:profiles_jhk}} 
\end{figure}

Launhardt et al. (2002) estimate the extinction due to interstellar dust in the Galactic 
Disk/Bulge and due to the Nuclear Bulge to be of the same order ($A_{\rm V}=15$~mag). 
The total extinction
to the Galactic Center is thus $A_{\rm V}=30$~mag. Due to the $\lambda ^{2}$ law of
the dust opacity, the stars in the Nuclear Bulge are fainter in the J and H band.
Thus, these bands are more contaminated by Galactic Bulge stars than the K$_{\rm S}$
band. This can be qualitatively seen in Fig.~\ref{fig:profiles_jhk} that shows a cut
along Galactic longitude through Sgr~A$^{*}$. The central stellar cluster,
which is most prominent in the K$_{\rm S}$ band, is more and more buried in the continuum 
due to foreground stars in the J and H band. 
Since we are only interested in the stellar distribution of the Nuclear Bulge, we
only use the K$_{\rm S}$ band image and the K$_{\rm S}$--H color for our studies.

\begin{figure}
	\psfig{file=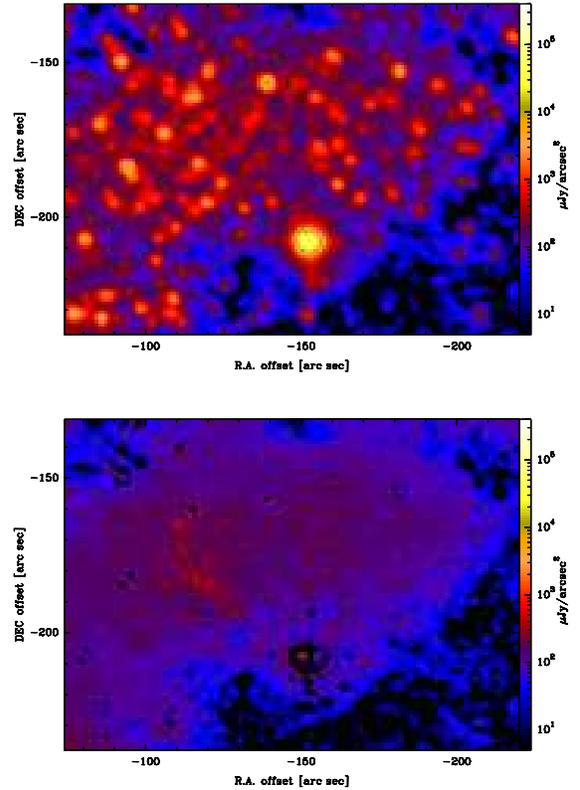,angle=-90,width=8cm}
      \caption{Upper panel: a subfield of the original K$_{\rm S}$ band image. 
	Lower panel: same image after subtraction of Lorentzian fits to the point sources.
	The offsets are given with respect to Sgr~A$^{*}$. 
	\label{fig:subtract}} 
\end{figure}

In both bands, K$_{\rm S}$ and H, single, distinguishable point sources are 
visible. The point spread function of the 2MASS data does not have the shape 
of the seeing-determined point spread function and is difficult to approximate by an analytic formula.
We use a modified Lorentzian profile to fit and subtract the distinct point sources.
The usage and advantages of a modified Lorentzian profile are described in Philipp et al. (1999).
Fig.~\ref{fig:subtract} shows an example of a small field before and after subtracting the
point sources. 

\begin{figure}
	\psfig{file=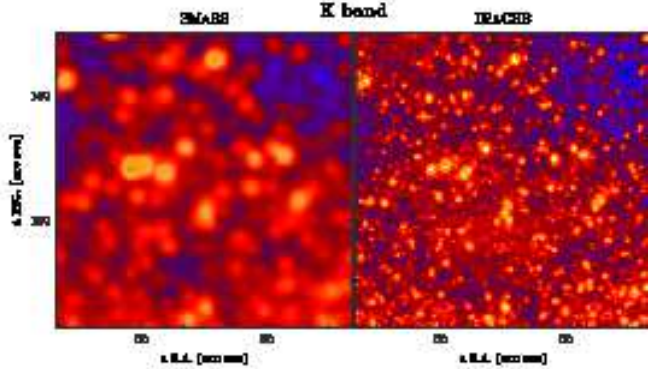,angle=-90,width=\hsize}
      \caption{K$_{\rm S}$ band images of a small subfield $\sim$5~pc north-east from Sgr~A$^{*}$.
	Left panel: 2MASS image with 3$''$ effective resolution. Right panel: IRAC2B 
	image with 1$''$ resolution (Philipp et al. 1999). The coordinates are offsets with respect to
	Sgr~A$^{*}$.
	\label{fig:stars}} 
\end{figure}

Since the effective resolution of 3$''$ is not sufficient to resolve all stars,
there are often several in projection closely packed stars that appear as a point source in
the 2MASS image. In order to illustrate this effect, we show in 
Fig.~\ref{fig:stars} a small area $\sim$5~pc north-east of Sgr~A$^{*}$
of the 2MASS image together with an image of the same region observed with IRAC2B
(Philipp et al. 1999). 

The IRAC2B image has a seeing of of $\sim$1$''$. Clearly, several point sources in the 2MASS 
image are resolved into multiple point sources in the IRAC2B image. This effect complicates
the shape of the 2MASS point source profiles and makes a subtraction difficult.
Our point source detection limit is $\sim$0.3~mJy, the completeness limit is $\sim$3~mJy
in the K$_{\rm S}$ band. In this way we found $\sim$75\,000 point sources in the whole
field and $\sim$13\,500 point sources in our region of interest (see Fig.~\ref{fig:mosaic}).

\begin{figure}
	\resizebox{\hsize}{!}{\includegraphics{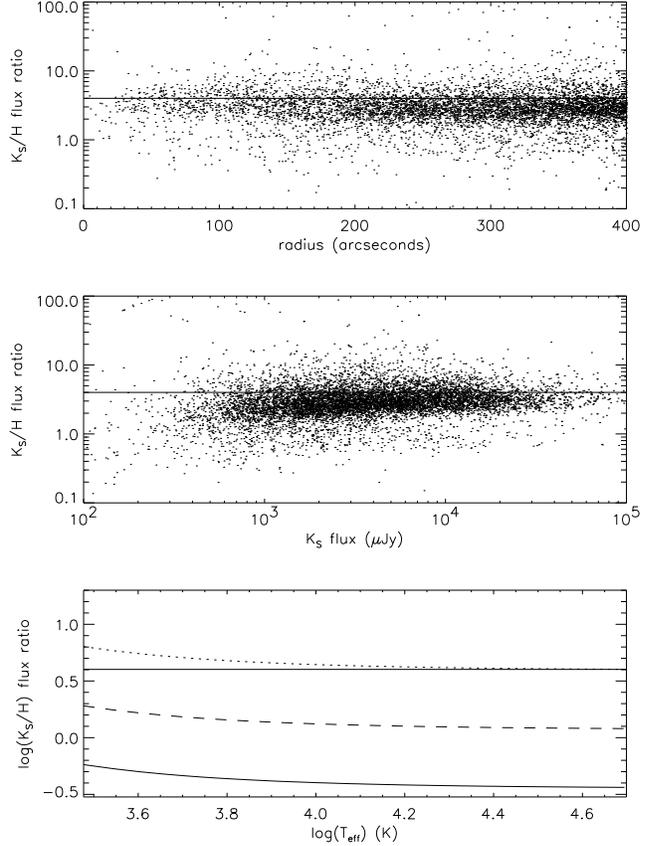}}
      \caption{Upper panel: K$_{\rm S}$/H flux ratio distribution 
	of the point sources as a function of their distance to Sgr~A$^{*}$. Middle panel:
	K$_{\rm S}$/H flux ratio distribution as a function of the K$_{\rm S}$ band flux.
	Lower panel: theoretical K$_{\rm S}$/H flux ratio for stars with different
	effective temperatures. Solid line: no reddening. Dashed line: reddening
	corresponding to $A_{\rm V}$=15~mag. Dotted line: reddening
	corresponding to $A_{\rm V}$=30~mag.
	\label{fig:starshk}} 
\end{figure}

For the color determination we sum the flux of all point sources
in a circle of 3$''$ (the seeing) diameter
around the positions of a given point source in the K$_{\rm S}$ band.
In a second step, we sum the flux of all point sources in the H band within
a circle of the same diameter. This proved to be the best method for an
accurate correlation of a sufficient number of point sources in both bands.
In this way we find $\sim$60\,000 point sources in both bands in the whole field and
$\sim$12\,000 point sources in our region of interest (see Fig.~\ref{fig:mosaic}).
Fig.~\ref{fig:starshk} shows the K$_{\rm S}$/H flux ratio distribution 
of the point sources as a 
function of their distance to Sgr~A$^{*}$ (upper panel) and the K$_{\rm S}$/H flux ratio 
distribution as a function of the K$_{\rm S}$ band flux (middle panel). 
The point sources clearly become redder towards the Galactic Center. The trend that
faint sources are bluer is due to our way of finding correlated point sources
and is not a physical effect. In order to discriminate between the stars in the
Nuclear Stellar Disk and the Nuclear Stellar Cluster, we use a limiting K$_{\rm S}$/H flux
ratio of 4, which is represented as a horizontal line in Fig.~\ref{fig:starshk}.
This limit is chosen in a way to assign the majority of the stars to the
Nuclear Stellar Disk. The remaining stars belong with a high probability
to the Nuclear Stellar Cluster. 

\begin{figure}
	\resizebox{\hsize}{!}{\includegraphics{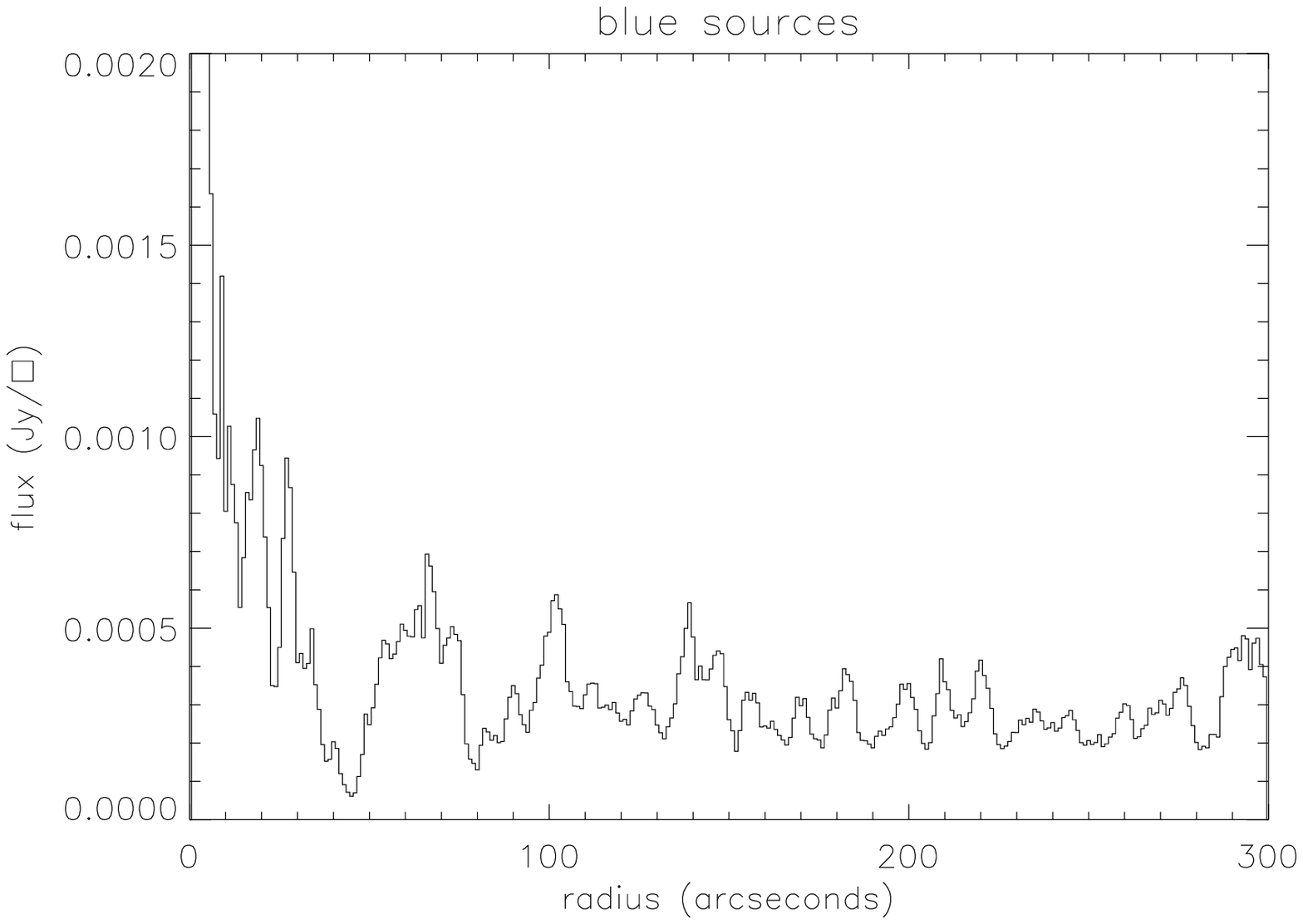}}
      \caption{Radial profile of the blue point sources averaged over ellipsoids 
	with an axis ratio of 1.4:1. \label{fig:bluestars}} 
\end{figure}

Fig.~\ref{fig:bluestars} shows the radial distribution of the K$_{\rm S}$ band
flux of the blue point sources averaged over ellipsoids with an axis ratio of 1.4:1,
which is consistent with the K$_{\rm S}$ isocontours of the Nuclear Bulge.
From 100$''$ to 300$''$ the radial distribution is almost flat and has a value of 
$\sim$300~$\mu$Jy.
The standard deviation increases with decreasing radius, because of the decreasing area
of the rings of integration near the center. The inner maximum is mainly due to one 
luminous point source, which is the central cluster of He{\sc i} stars.
There is also a slight increase of the mean integrated flux density
to the center to a value of $\sim$400~$\mu$Jy. Since the K$_{\rm S}$ band intensity
reaches maximum values of $\sim$10~mJy the radial profile can be regarded as constant even
for small radii. We conclude that the blue point sources represent an almost homogeneous
foreground distribution that can be removed without changing the distribution of the
Nuclear Stellar Cluster.

With our K$_{\rm S}$ band completeness limit of $\sim$3~mJy we can observe
hot main sequence stars, cold giants, hot and cold supergiants, and Wolf-Rayet
stars as distinguishable point sources (see Fig.~D1 of Philipp et al. 1999). 
The lower panel of Fig.~\ref{fig:starshk}
shows the theoretically expected K$_{\rm S}$/H band flux ratios for stars
with a given effective temperature (i) without reddening, (ii) with a reddening
corresponding to $A_{\rm V}$=15~mag (Galactic Disk/Bulge), and (iii)  with a reddening
corresponding to $A_{\rm V}$=30~mag (Galactic Disk/Bulge and Nuclear Bulge).
We use the dust opacities of Launhardt et al. (2002) and
$A_{\rm V}/N_{\rm H}=3.25\,10^{-22}$, which is a factor $\sim$1.6 lower than
the value given by the same authors. With a total reddening of $A_{\rm V}$=30~mag
and a limiting K$_{\rm S}$/H flux ratio of 4 we theoretically expect to find
mainly supergiants of temperatures smaller than 30\,000~K. The fact that
we classify the central He{\sc i} star cluster as a blue point source justifies 
a posteriori the $A_{\rm V}/N_{\rm H}$ we use. 

\begin{figure}
	\resizebox{\hsize}{!}{\includegraphics{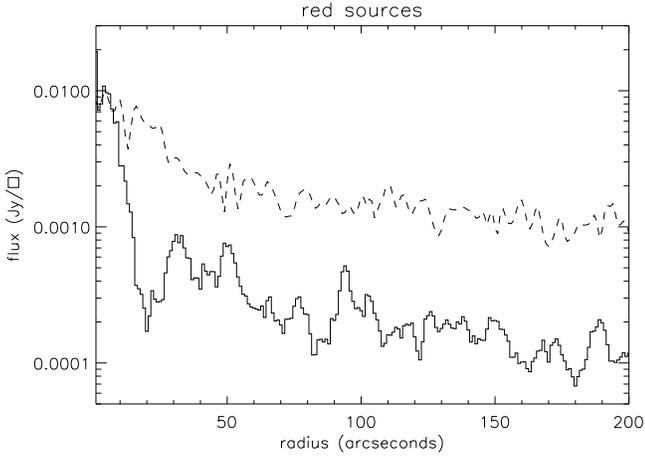}}
      \caption{Solid line: radial profile of the red point sources
	averaged over ellipsoids with an axis ratio of 1.4:1.
	Dashed line: Slice parallel to the galactic longitude through Sgr~A$^{*}$
	through the K$_{\rm S}$ image where all point sources (red and blue) are subtracted.	
	\label{fig:redstars}} 
\end{figure}

Fig.~\ref{fig:redstars} shows the radial distribution of the K$_{\rm S}$ band
flux of the red point sources averaged over the same ellipsoids.
The flux of this stellar population is dominated by sources of flux densities between 10 and 100~mJy.
Its radial profile is markedly different from that of the blue point sources. It rises steeply
in the inner 50$''$ from 500~$\mu$Jy to 10~mJy in the center. For $R > 20''$ the flux
density of the red point sources represents only a small fraction ($<20$\%) of the flux density
of the unresolved background (Fig.~\ref{fig:redstars}). Thus, the red point sources represent 
Central Stellar Disk stars, which justifies our limiting K$_{\rm S}$/H  flux ratio.

The total flux density of the blue point sources is 155~Jy, that of the red point sources
110~Jy. We find a flux density of the background of 335~Jy. Philipp et al. (1999)
analyzed K band data of a $10' \times 10'$ field centered on Sgr~A$^{*}$.
They found a total K band flux of 752~Jy and
a total stellar flux density of 370~Jy and estimated the flux of the 
background to be 283~Jy. Their total flux density (blue and red sources included) 
is 25\% higher than that of the 2MASS image ($S$=600~Jy). The fraction of the flux
density of fitted point sources to the total flux density is 0.44 for the 2MASS data and 0.49
for the data of Philipp et al. (1999). Thus, we conclude that our fitting procedure
works in a satisfactory way.

In the end we made two different images:
\begin{enumerate}
\item
an image where only the blue foreground point sources are subtracted,
\item
an image where all point sources that could be identified are subtracted.
\end{enumerate}

\section{Results \label{sec:results}}

\begin{figure}
	\resizebox{\hsize}{!}{\includegraphics{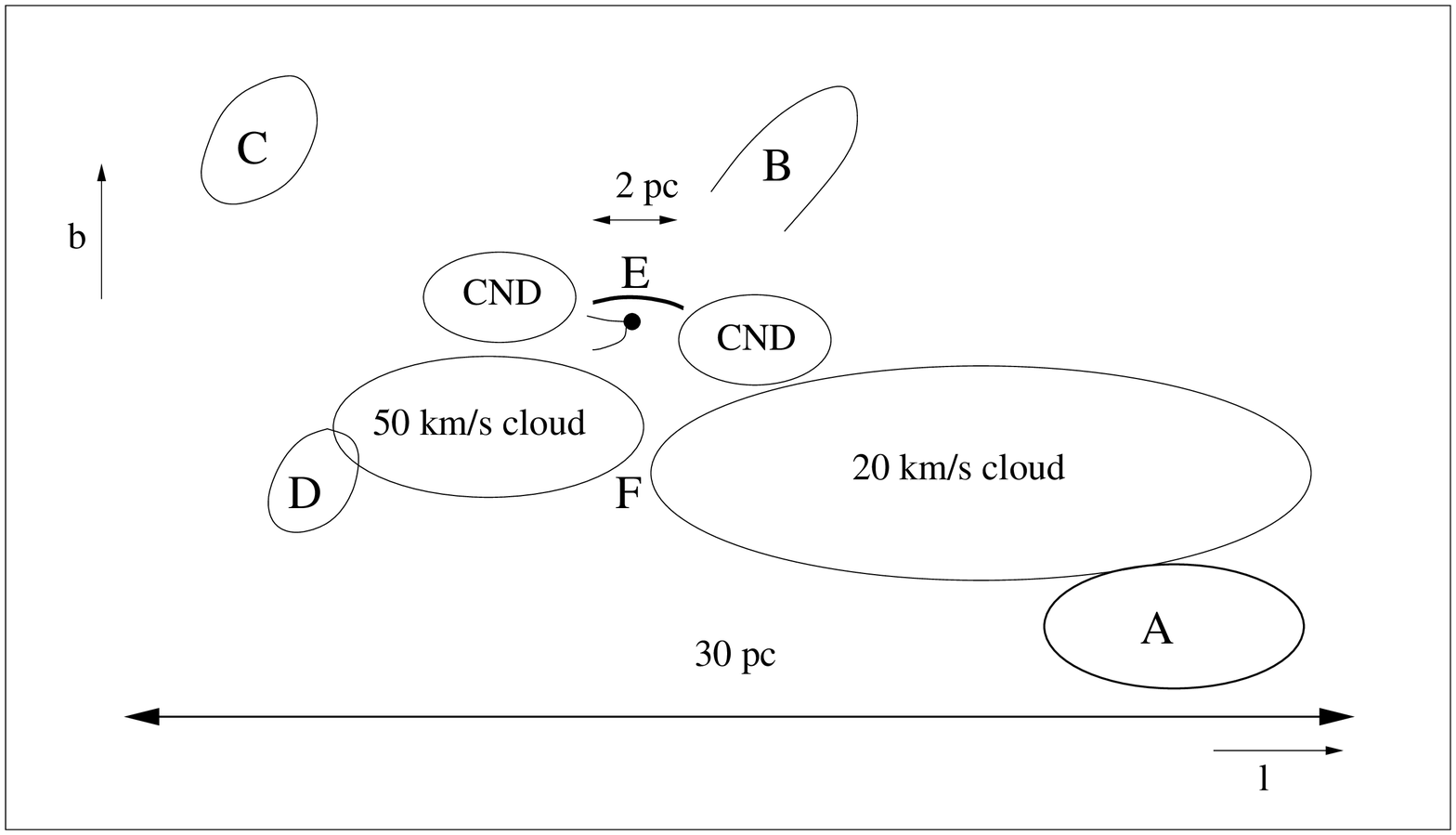}}
      \caption{Sketch of the inner 30~pc of the Galaxy. The central black dot represents the
	He{\sc i} star cluster surrounded by the Minispiral and the CND (where only the
	high velocity lobes are shown). The 20 and 50~km\,s$^{-1}$ giant molecular clouds
	are located at negative latitudes. Capital letters indicate structures
	discussed in the text. 1~pc corresponds to 24$''$.
	\label{fig:gcsketch}} 
\end{figure}

In the inner 100~pc of the Galaxy the gas is highly clumped in giant molecular clouds with
a volume filling factor of $\sim$1\% (Launhardt 2002). 
Following Zylka et al. (1990) three main giant molecular cloud complexes can be distinguished: 
(i) Sgr~A East Core, (ii) the $\sim$50~km\,s$^{-1}$ cloud, and (iii) the 20~km\,s$^{-1}$ cloud.
Sgr East core is part of the 50~km\,s$^{-1}$ cloud complex, thus we will treat these
features as a single structure. Fig.~\ref{fig:gcsketch} shows a sketch of the projected 
inner 30~pc of the Galaxy, where the main features are indicated 
(Minispiral, CND, 20~km\,s$^{-1}$ cloud, 50~km\,s$^{-1}$ cloud).

\begin{figure*}
	\psfig{file=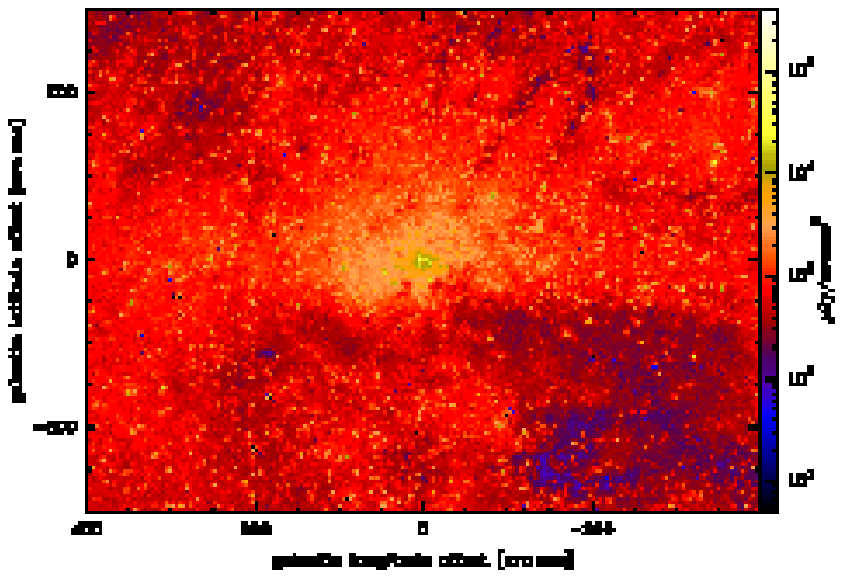,angle=-90,width=12cm}
	\psfig{file=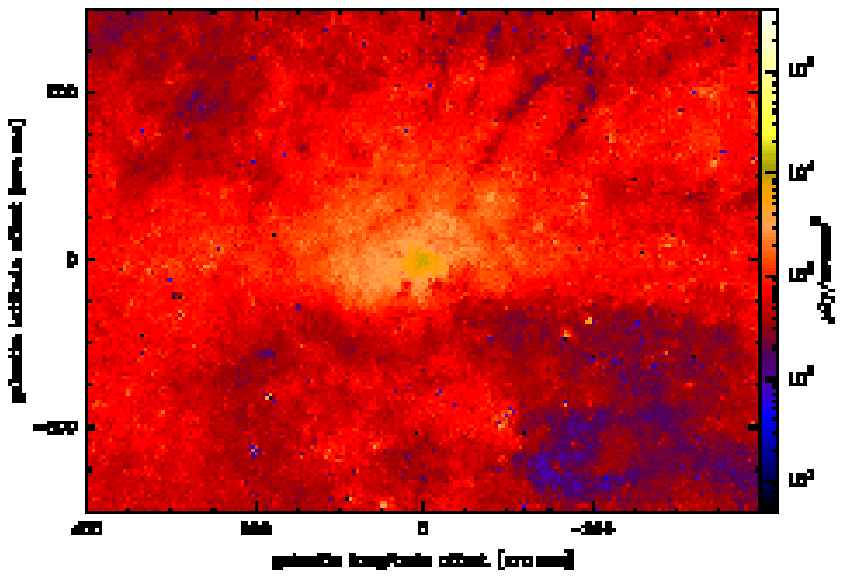,angle=-90,width=12cm}
      \caption{K$_{\rm S}$ band image of the Galactic Center region.
Upper panel: only the blue point sources are subtracted. 
Lower panel: all distinguishable point sources are subtracted. \label{fig:kband}} 
\end{figure*}

Fig.~\ref{fig:kband} shows the K$_{\rm S}$ band image of the Galactic Center region
(i) with only the blue point sources subtracted, (ii) with all distinguishable
point sources subtracted. 
In both images the absorption features are very similar. This shows again that the foreground stars,
despite their extended wings of their profiles, do not affect considerably
the underlying continuum emission. Thus, the image where all point sources (blue and red) 
are subtracted has a less discrete character, i.e. it is less noisy.

In the inner $10''$ around Sgr~A$^{*}$,
i.e. in the very central star cluster, the procedure of point source subtraction
is not reliable. It mainly subtracted the central peak. Thus, this region
has to be discarded in the discussion using the image where all point sources
are subtracted. 
In the following we will discuss the main features of both images, because of 
their similarity.

One clearly can distinguish the 20~km\,s$^{-1}$ and the 50~km\,s$^{-1}$ cloud
complexes at negative galactic latitudes (see Fig.~\ref{fig:gcsketch}). 
The 20~km\,s$^{-1}$ cloud complex has
an almost linear edge to positive galactic latitudes. It covers a region of
nearly 7$'$ in galactic longitude. The depth of the absorption is almost 
constant from $\Delta l \sim -80''$ to $\Delta l \sim -400''$ with respect to
Sgr~A$^{*}$. The 50~km\,s$^{-1}$ cloud complex is separated into two components.
The first, located below Sgr~A$^{*}$ at negative galactic latitudes, is curved.
The second component runs almost perpendicular to the first at a galactic
longitude of $\Delta l \sim 200''$ with respect to Sgr~A$^{*}$ (D in Fig.~\ref{fig:gcsketch}). 
Whereas the absorption caused by the first component is almost uniform, the second component
shows a gradient. The absorption depth of the first component is smaller than
that of the 20~km\,s$^{-1}$ cloud complex.

Furthermore, there is a large cloud
complex at $\Delta l \sim -200''$, $\Delta b \sim -250''$
that forms a shell-like structure (A in Fig.~\ref{fig:gcsketch}). 
The part towards Sgr~A$^{*}$ has a larger absorption depth than the opposite side.

At positive galactic latitudes one can find two further absorption features. One
at positive galactic longitudes that appears to be elongated diagonal to
the image axis (C in Fig.~\ref{fig:gcsketch}). 
The second one at negative galactic longitudes has the shape of the 
letter U (B in Fig.~\ref{fig:gcsketch}). 
We believe that this is a foreground cloud and we will not discuss it further.

It is important to note that there is almost no absorption along the galactic plane 
and little absorption for positive galactic latitudes at $\Delta l =0$ with
respect to Sgr~A$^{*}$.

\section{Reconstruction of the LOS distribution \label{sec:reconstruct}}

In this Section we describe the method to calculate the distance along the 
line--of--sight (LOS) of the molecular cloud complexes observed in 
K$_{\rm S}$ band absorption. We model the K$_{\rm S}$ band continuum emission
distribution using analytical expressions for the stellar volume emissivity.
We use the image where all point sources are subtracted, but we also apply the method to
the image where only the red point sources are subtracted to make sure that
the subtraction of the red point sources does not alter the results.

For the reconstruction we make three basic assumptions:
\begin{enumerate}
\item
We assume an axis--symmetric distribution of stars in the central 7$'$ around
Sgr~A$^{*}$.
\item
We assume that the clouds are optically thick and that their area coverage factor 
is $\sim$1. This means that the clouds block the light of all stars located
behind them. The average column density of the inner 100~pc of the Galaxy is 
$\sim$7\,10$^{22}$~cm$^{-2}$ (Launhardt et al. 2002). 
Assuming an area filling factor of 10\% the mean density
is 10$^{4}$~cm$^{-3}$, which is about the critical density to resist tidal shear
(Vollmer \& Duschl 2001).
Thus the column density of a giant molecular cloud is $\sim$7\,10$^{23}$~cm$^{-2}$
leading to a K band extinction of $A_{\rm K} \sim 28$~mag (using $A_{\rm K}/A_{\rm V}=0.122$,
Mathis et al. 1983).
\item
We assume a homogeneous large scale reddening gas layer within the Nuclear Bulge
in front of the GMCs, which is due to the pervasive, low density phase of the ISM.
For a field twice as large as our region of interest, we observe a variation in the
large scale reddening.
\end{enumerate}

\subsection{The method \label{sec:method}}

We first fit an analytic profile to the volume emissivity of the central stellar
distribution in the inner 60~pc around Sgr~A$^{*}$. Following Launhardt et al. (2002)
we use two components: (i) the Nuclear Stellar Cluster at $R < 1$~pc and (ii)
the Nuclear Stellar Disk for radii $R < 100$~pc around Sgr~A$^{*}$.
We use the following analytic expressions: 
\begin{equation}
({\rm i})\ \ \rho_{1}(x,y,z)=\frac{\rho_{0}}{1+(\frac{x}{r_{0}})^{2}+(\frac{y}{r_{1}})^{2}+
(\frac{z}{r_{0}})^{2}}\ ,
\end{equation}
\begin{equation}
({\rm ii})\ \ \rho_{2}(x,y,z)=\frac{\rho_{1}}{\sqrt{(\frac{x}{r_{2}})^{2}+
(\frac{z}{r_{2}})^{2}}}\times {\rm sech}^{2}\big(\frac{y}{y_{0}}\big)\ ,\  {\rm and\ alternatively}
\label{eq:alt1}
\end{equation}
\begin{equation}
\ \ \ \ \ \ \rho_{2}(x,y,z)=\tilde{\rho}_{1}\,\exp\big({\rm ln}(\frac{1}{2})|R/c|\big)
\times {\rm sech}^{2}\big(\frac{y}{y_{0}}\big)\ ,
\label{eq:alt2}
\end{equation}
where $x$/$y$ are the distances along the galactic longitude/latitude, $z$ is the 
distance along the LOS with respect to Sgr~A$^{*}$, $R=\sqrt{x^{2}+y^{2}+z^{2}}$,
$\rho_{0} / \rho_{1} = 5\,10^{3}$, $\rho_{0} / \tilde{\rho}_{1}=6.6\,10^{4}$,
$r_{0}$=0.1~pc, $r_{1}$=0.07~pc, $r_{2}$=1~pc, $y_{0}$=10~pc, and $c$=120~pc.
The profile of Eq.~(\ref{eq:alt2}) was designed to fit the COBE
data of 0.7$^{\rm o}$ resolution (Launhardt et al. 2002). Since a constant radial
density is dynamically difficult to explain, we give an alternative profile
of the form $\rho_{2} \propto R^{-1}$ (Eq.~(\ref{eq:alt1})).
The total volume emissivity is $\rho(x,y,z)=\rho_{1}(x,y,z)+\rho_{2}(x,y,z)$ in model units.

\begin{figure}
	\resizebox{\hsize}{!}{\includegraphics{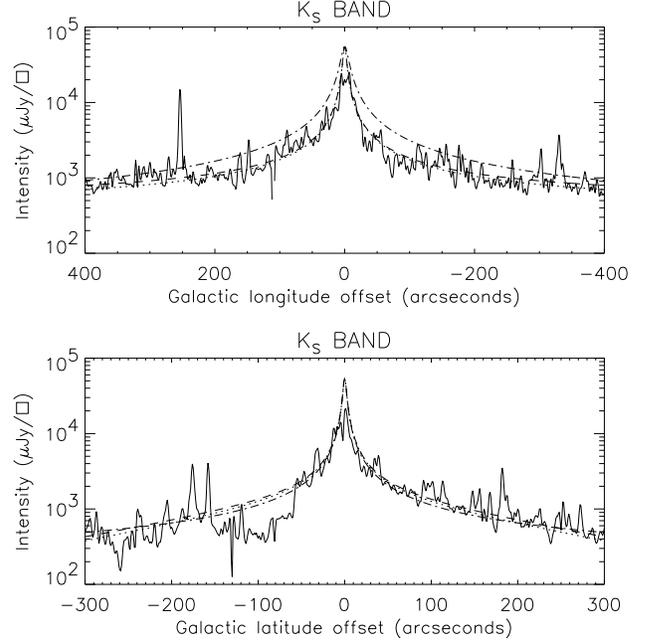}}
      \caption{Slices of the model intensities along the galactic longitude and latitude
	through Sgr~A$^{*}$ together with slices of the K$_{\rm S}$ data.
	Dotted line: Eq.~(\ref{eq:alt1})/(\ref{eq:inti}) $a=0$, $z_{0}=$-50~pc. Dashed line:  
	Eq.~(\ref{eq:alt1})/(\ref{eq:inti})
	$a=0$, $z_{0}=-100$~pc. Dash-dotted line: Eq.~(\ref{eq:alt2})/(\ref{eq:inti}) $a=0$.
	1~pc corresponds to 24$''$. \label{fig:kprofiles}} 
\end{figure}

The volume emissivity is integrated along the LOS.
As a next step a constant offset is subtracted from the K$_{\rm S}$ band data to account for
homogeneous foreground emission. Then, the model map is multiplied by a factor 
$\xi$ to fit the K$_{\rm S}$ band image. 
This factor is found in minimizing the difference between the model and the
observed data. The result of this procedure can be seen in Fig.~\ref{fig:kprofiles},
where slices of the resulting model intensity along the galactic longitude and latitude
through Sgr~A$^{*}$ together with slices of the K$_{\rm S}$ data are shown.
The model slices of Eq.~(\ref{eq:alt1}) nicely fit the data in the regions without absorption.
Since the model distribution along the Galactic latitude has the same form for
Eq.~(\ref{eq:alt1}) and Eq.~(\ref{eq:alt2}), it fits equally well the observed K$_{\rm S}$ band
emission distribution. The emission distribution along the Galactic Longitude is overestimated
by Eq.~(\ref{eq:alt2}). We conclude that Eq.~(\ref{eq:alt1}) fits the data better.

We define 
\begin{equation}
I_{0}(x,y)=\xi \int^{z_{1}}_{z_{0}} \rho (x,y,\tilde{z}) {\rm d}\tilde{z}\ ,
\label{eq:inti}
\end{equation}
\begin{equation}
I_{z}(x,y)=\xi \int^{z}_{z_{0}} \rho (x,y,\tilde{z}) {\rm d}\tilde{z}\ .
\end{equation}
where the lower boundary $z_{0}$ is negative and fixed and $z_{1}=|z_{0}|$. 
Let $I_{\rm K}(x,y)$ be the K$_{\rm S}$ band intensity. 
Then, the LOS distance $z$ for a given position $(x,y)$ is given by
\begin{equation}
\frac{I_{z}(x,y)}{I_{0}(x,y)} = \frac{I_{\rm K}(x,y)}{I_{0}(x,y)}\ ,
\end{equation}
or 
\begin{equation}
I_{z}(x,y) = I_{\rm K}(x,y)\ .
\end{equation}

Since the intensity of sky is not known, we will generalize our method in
subtracting a constant intensity $a$ from the image. The sky intensity of
Fig.~\ref{fig:kband} is determined by assuming that the darkest region in the
whole field (cf. Fig.~\ref{fig:mosaic}) has zero intensity. Thus, the equation for the
determination of the LOS distance of gas located at ($x$, $y$) yields:
\begin{equation}
I_{\rm K}(x,y) - a = \xi I_{z}(x,y)\ ,
\label{eq:offset}
\end{equation} 
In order to investigate how $z$ varies with $\xi$ and $a$ we set 
\begin{equation}
I_{z}(x,y) = \int_{z_{0}}^{z} \rho_{1}(x,y,\tilde{z}) {\rm d}\tilde{z} =
\xi^{-1}(I_{\rm K}(x,y)-a)\ .
\end{equation}
Solving this equation for $z$ yields
\begin{equation}
\tilde{z} = (1+x^{2}+y^{2}) \tan \big( (1+x^{2}+y^{2})(\rho_{0}^{-1}(\xi^{-1}(I_{\rm K}(x,y)-a))) -
\frac{\pi}{2} \big)\ .
\end{equation}
thus, small variations in $I_{\rm K}(x,y)$ lead to large variations in $z$ for
large projected distances and small $I_{\rm K}(x,y)$, i.e. large LOS distances $z$.

\begin{figure}
	\resizebox{\hsize}{!}{\includegraphics{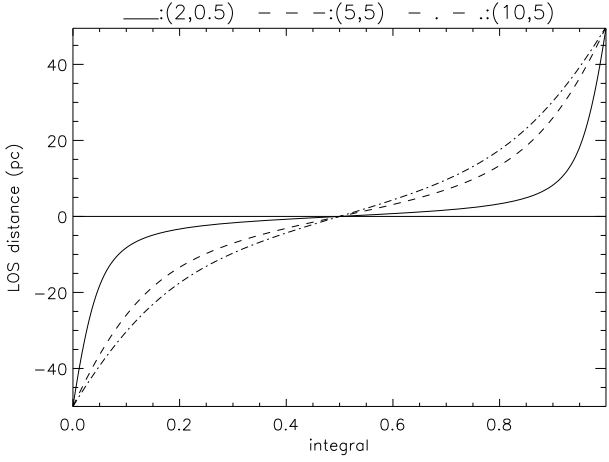}}
      \caption{LOS distance $z$ as a function of the ratio
	$I_{z}/I_{0}$ for 3 different projected distances:
	Solid line: (2~pc, 0.5~pc), dashed line: (5~pc, 5~pc), dot-dashed line:
	(10~pc, 5~pc). \label{fig:LOS_int}} 
\end{figure}

In order to illustrate this effect for the realistic volume emissivity $\rho=\rho_{1}+\rho_{2}$,
we show $z$ as a function of $I_{z}/I_{0}$ for 3 different projected distances
in Fig.~\ref{fig:LOS_int}.
For deep absorption features small variations in $I_{z}/I_{0}$ lead to large
variations in $\tilde{z}$. The non-linear regime begins at smaller LOS distances
for large projected distances. The error of the LOS distance for large
projected and LOS distances can be of the order of $\sim$30\%.

Due to the radial distribution of the volume emissivity, this method detects
easily clouds in front of Sgr~A$^{*}$. Clouds that are located behind 
Sgr~A$^{*}$ show only small absorption features that can be buried by
the discrete character of the signal (single stars).

Clearly, the LOS distance depends strongly on the offset $a$ (Eq.~(\ref{eq:offset}))
applied on the data. There is no way to determine a priori this constant.
In addition, due to the form of $\rho_{2}$ in Eq.~(\ref{eq:alt1})
the LOS distance depends on $z_{0}$.

We made 4 different calculations for the profile of Eq.~(ref{eq:alt1}) and
2 calculations for the profile of Eq.~(\ref{eq:alt2}) to take these effects into account.
For Eq.~(\ref{eq:alt1}) we set:
\begin{enumerate}
\item
$z_{0}$=-50~pc, $a$=0;
\item
$z_{0}$=-50~pc, $a$=150~$\mu$Jy/arcsec$^{2}$;
\item
$z_{0}$=-100~pc, $a$=0;
\item
$z_{0}$=-100~pc, $a$=150~$\mu$Jy/arcsec$^{2}$;
\end{enumerate}
for Eq.~(\ref{eq:alt2}) we set:
\begin{enumerate}
\item
$z_{0}$=-150~pc, $a$=0;
\item
$z_{0}$=-150~pc, $a$=150~$\mu$Jy/arcsec$^{2}$.
\end{enumerate}
Since the profile of Eq.~(\ref{eq:alt2}) has a cutoff at $R$=120~pc, it is not
necessary to vary the the lower integration limit $z_{0}$ of Eq.~(\ref{eq:inti}).

The value of the offset $a$ is chosen such that the deepest absorption at $\Delta l \sim -200''$,
$\Delta b \sim -200''$, which most probably belongs to the 20~km\,s$^{-1}$ cloud complex, is 
close to zero (cf. Sect.~\ref{sec:rrr}). For comparison, Launhardt et al. (2002) estimate the 
K band flux of the Galactic Disk and Bulge to be $\sim$20~MJy/sr=470$\mu$Jy/arcsec$^{2}$ and that
of the COBE peak emission of the Nuclear Bulge to be $\sim$10~MJy/sr=235$\mu$Jy/arcsec$^{2}$.
This implies that cloud A (Fig.~\ref{fig:gcsketch}) is located within the Nuclear Bulge.

\subsection{Results \label{sec:rrr}}

\begin{figure}
	\resizebox{\hsize}{!}{\includegraphics{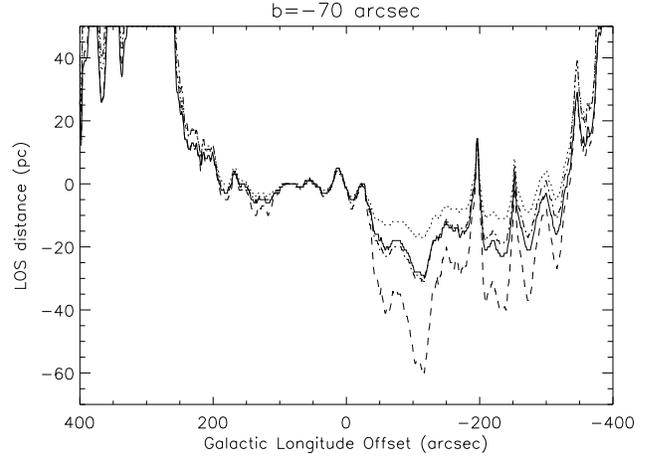}}
	\caption{Reconstruction of the LOS distances for a slice at
	$\Delta b = -70''$ parallel to the galactic longitude
	using Eq.~(\ref{eq:alt1}).
	Solid line: $a$=150~$\mu$Jy/arcsec$^{2}$ and $z_{0}=-50$~pc.
	Dotted line: $a$=0 $z_{0}=-50$~pc. Dashed line:
	$a$=150~$\mu$Jy/arcsec$^{2}$ and $z_{0}=-100$~pc.
	Dot-dashed line: $a$=0 $z_{0}=-100$~pc. Negative distances are
	in front of Sgr~A$^{*}$. \label{fig:LOSprofiles}}
\end{figure}

In order to demonstrate the differences between the 4 different reconstructions
using Eq.~(\ref{eq:alt1}),
we show in Fig.~\ref{fig:LOSprofiles} a slice through the reconstructed map at 
$\Delta b = -70''$ parallel to the galactic longitude. This is a representative cut
through the 50 and 20~km\,s$^{-1}$ cloud complexes. For clarity we applied
a median filter with a size of 11 pixels to the map. Negative distances are
in front Sgr~A$^{*}$.
As expected, the larger $|z_{0}|$ and $a$ the smaller is the LOS distance of
the clouds, i.e. the nearer it is to the observer. For the deepest absorption
at $\Delta l = -100''$ the LOS distance is $d \sim-60$~pc for $z_{0}=-100$~pc, 
$a$=150~$\mu$Jy/arcsec$^{2}$ and only $d \sim-17$~pc for $z_{0}=-50$~pc, $a$=0.
With increasing $z_{0}$ and $a$ the variation of the LOS distance within the
20~km\,s$^{-1}$ increases rapidly.
On the other hand, the LOS distance of the 50~km\,s$^{-1}$ cloud complex
($0'' < \Delta l < 200''$) does not vary much with varying $z_{0}$ and $a$, because it is 
located very close to Sgr~A$^{*}$ (0~pc$< d <-5$~pc).

\begin{figure}
	\resizebox{\hsize}{!}{\includegraphics{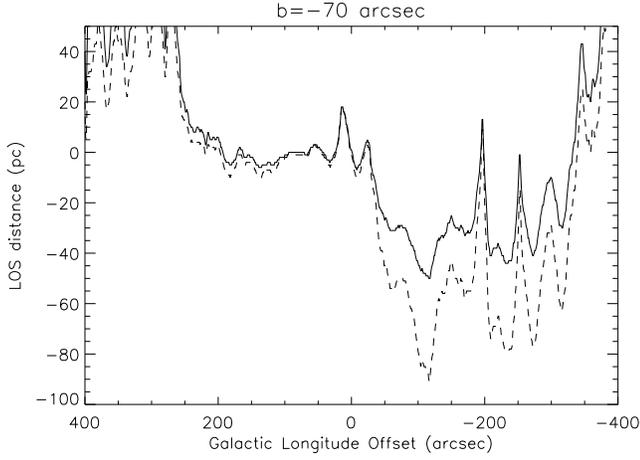}}
	\caption{Reconstruction of the LOS distances for a slice at
	$\Delta b = -70''$ parallel to the galactic longitude using Eq.~(\ref{eq:alt2}).
	Solid line: $a$=0 and $z_{0}=-150$~pc. Dashed line:
	$a$=150~$\mu$Jy/arcsec$^{2}$ and $z_{0}=-150$~pc.
	Negative distances are in front of Sgr~A$^{*}$. \label{fig:losalt2}}
\end{figure}

Fig.~\ref{fig:losalt2} shows the 2 different reconstructions using Eq.~(\ref{eq:alt2}).
The LOS distances of the 50~km\,s$^{-1}$ cloud complex are comparable to those 
calculated using Eq.~(\ref{eq:alt1}). However, the LOS distances from Sgr~A$^{*}$
of the 20~km\,s$^{-1}$ cloud complex are systematically larger than those calculated 
using Eq.~(\ref{eq:alt1}). This is due to the too high model continuum with respect to
the K$_{\rm S}$ band continuum emission (Fig.~\ref{fig:kprofiles}).
The relative distances between the molecular cloud complexes and the relative gradients
of the LOS distance within a cloud complex are the same for both model profiles.

We chose a profile using Eq.~(\ref{eq:alt1}) with $z_{0}=-50$~pc and 
$a$=150~$\mu$Jy/arcsec$^{2}$ for the final 
reconstruction of the LOS distance distribution. This represents a compromise
between a possible underestimation of the LOS distances due to an offset
in the K$_{\rm S}$ band data and too much variation of the LOS distances within
the 20~km\,s$^{-1}$ cloud complex. One has to bear in mind that LOS distances $d > -10$~pc
have a small error, whereas distances $d < -10$~pc can be up to a factor of 2
smaller than given in the final map.

\begin{figure*}
	\psfig{file=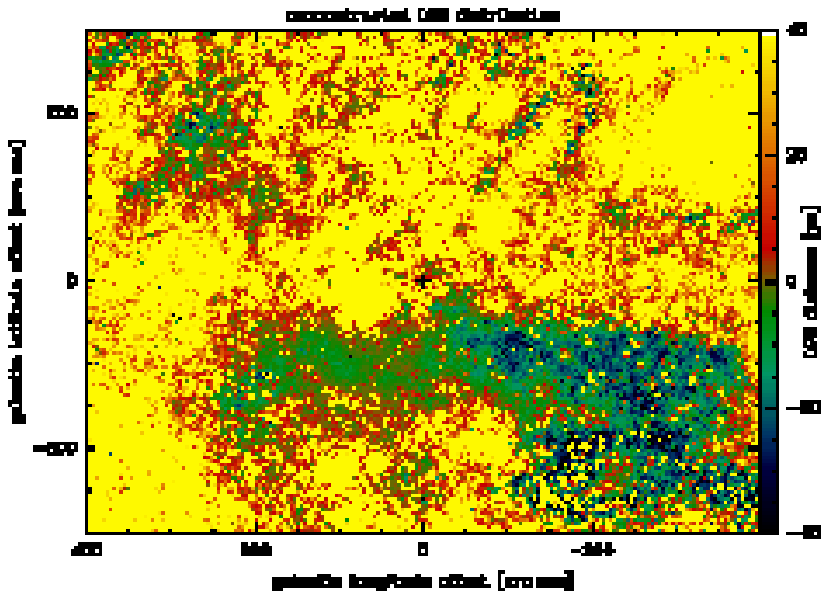,angle=-90,width=12cm}
	\psfig{file=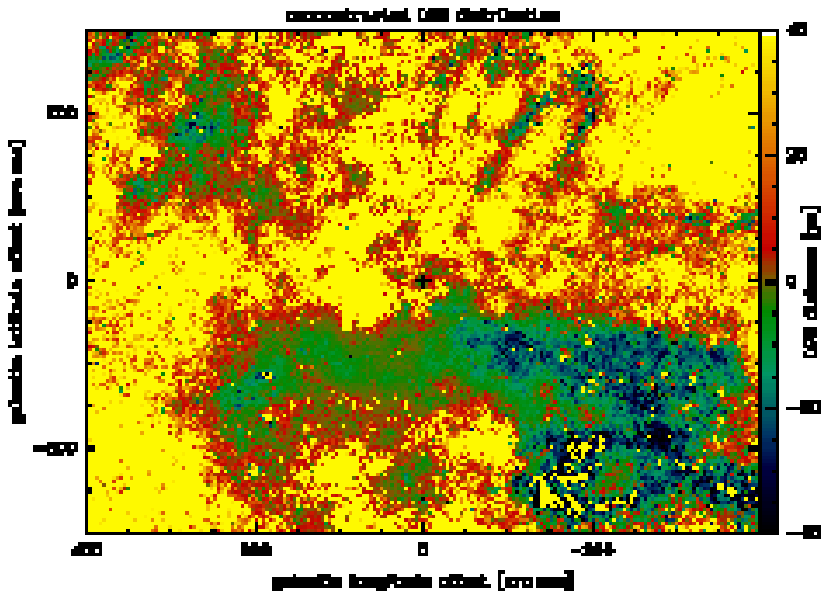,angle=-90,width=12cm}
      \caption{Map of the reconstructed LOS distances for the 2MASS K$_{\rm S}$
	images. Upper panel: all blue point sources are subtracted.
	Lower panel: all point sources (blue and red) are subtracted.\label{fig:recon}} 
\end{figure*}

\begin{figure}
	\psfig{file=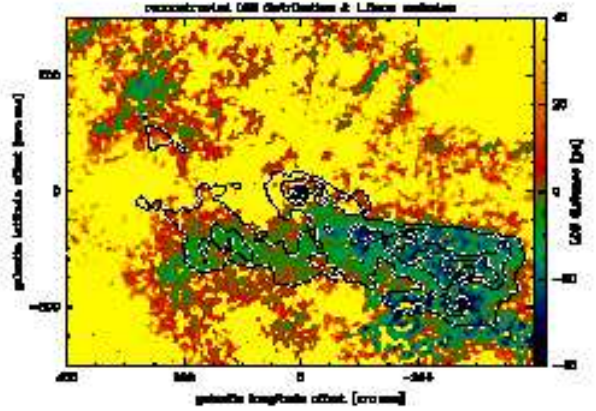,angle=-90,width=\hsize}
	\caption{Contours: IRAM 30m 1.2~mm observations of Zylka et al. (1998).
	Colors: LOS distance distribution filtered with a median filter of 11 pixels.
	\label{fig:overl_1.2mm}}
\end{figure}

\begin{figure}
	\psfig{file=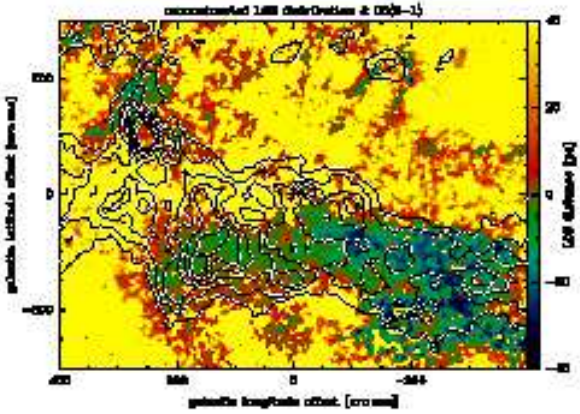,angle=-90,width=\hsize}
	\caption{Contours: IRAM 30m CS(2-1) line observations (G\"{u}sten et al. in prep.)
	integrated over the whole velocity range. 
	Colors: LOS distance distribution filtered with a median filter of 11 pixels.
	\label{fig:overl_cs2_1}}
\end{figure}

Fig.~\ref{fig:recon} shows the map of the reconstructed LOS distances for the
2MASS K$_{\rm S}$ images with $z_{0}=-50$~pc and $a$=150~$\mu$Jy/arcsec$^{2}$
for two cases: 
(i) all blue point sources are subtracted and (ii) all point sources (blue and red) 
are subtracted.
In order to correlate the reconstructed LOS distribution with observed clouds
we show in Fig.~\ref{fig:overl_1.2mm} the 1.2~mm observations of
Zylka et al. (1998) together with the LOS distance distribution filtered
with a median filter of 11 pixels.
As already shown by Philipp et al. (1999)
the 50~km\,s$^{-1}$ and the 20~km\,s$^{-1}$ cloud complexes can be clearly seen at
$200'' > \Delta l > 0''$ and $-50'' > \Delta l > -300''$, respectively.
The part of the 50~km\,s$^{-1}$ cloud near Sgr~A$^{*}$ is not seen in the
1.2~mm data, because of their observing mode (double beam mapping/chopping). 
The 50~km\,s$^{-1}$ cloud complex is located between 0~pc and 6~pc in front of Sgr~A$^{*}$
(cf. Fig.~\ref{fig:LOSprofiles}). We can identify a
small gradient of the LOS distance parallel to the galactic longitude.
From $\Delta l=150''$ to $\Delta l=50''$ the LOS distance increases from $-6$~pc to
0~pc. Then it drops again to $-4$~pc at $\Delta l \sim 20''$. This is best seen
in the lower image of Fig.~\ref{fig:recon}. The annex of the 50~km\,s$^{-1}$ cloud complex
that is located at ($\Delta l \sim 200''$, $\Delta b \sim -150''$) and which has an
elongation approximately perpendicular to the main 50~km\,s$^{-1}$ cloud complex is located
nearer to the observer ($d \sim-10$~pc) than the main 50~km\,s$^{-1}$ cloud complex.
We might tentatively see a gradient from large distances at small galactic
latitudes to small distances at larger galactic latitudes, i.e. near the main
50~km\,s$^{-1}$ cloud complex. This cloud is also not seen in the 1.2~mm observations, because
of the observing mode (double beam mapping/chopping). 
It clearly appears in the IRAM 30m CS(2-1) data 
(G\"{u}sten et al. in prep.; Fig.~\ref{fig:overl_cs2_1}).

The 20~km\,s$^{-1}$ cloud complex shows a very patchy structure compared to the
50~km\,s$^{-1}$ cloud complex. It is not excluded that both consist of several distinct
clouds. It has an overall LOS distance gradient from $\sim$0~pc
at $\Delta l = -350''$ to $\sim -27$~pc at $\Delta l = -120''$
(cf. Fig.~\ref{fig:LOSprofiles}). The gradient becomes
shallower for increasing $\Delta l$. For $-120'' < \Delta l < -50''$ the gradient
has the opposite sign, i.e. the 20~km\,s$^{-1}$ cloud complex approaches the Sgr~A$^{*}$.
We observe a jump of the LOS distance at $\Delta l \sim -30''$ which might represent
a discontinuity between the 20 and 50~km\,s$^{-1}$ cloud complexes, i.e. that both
structures are not physically connected (F in Fig.~\ref{fig:gcsketch}).

Below the 20~km\,s$^{-1}$ cloud complex a large shell-like absorption feature can be seen 
($\Delta l =-250''$, $\Delta b = -250''$) (A in Fig.~\ref{fig:gcsketch}). 
Since this structure does not appear
in the 1.2~mm data nor in the CS(2-1) data, it must be a structure
that is located distinnctly in front of the 20~km\,s$^{-1}$ cloud. 
In the maps of the reconstructed 
LOS distances it appears as a yellow region with blue borders. This means that
this absorption feature is located at $z < -50$~pc.
Since it is only this structure that shows negative absorption we are
confident that the applied offset $a$=150~$\mu$Jy/arcsec$^{2}$ is acceptable.
If $a$ was slightly larger, parts of the 20~km\,s$^{-1}$ cloud complex would have
negative absorption feature, which would not be acceptable.

At positive galactic longitudes and latitudes an absorption feature shows up that
is also located in front of Sgr~A$^{*}$ (C in Fig.~\ref{fig:gcsketch}). 
A comparison with the CS(2-1)
data (Fig.~\ref{fig:overl_cs2_1}) shows this cloud complex is really located 
in the Nuclear Bulge near Sgr~A$^{*}$.
At negative galactic longitudes and positive latitudes another shell-like structure
can be seen (B in Fig.~\ref{fig:gcsketch}). 
If it was located at distances $<$50~pc to Sgr~A$^{*}$ it would be
stretched within one rotation period, i.e. $<1$~Myr parallel to the galactic longitude, 
because of the strong shear motions in this region.
Thus, we believe that it is most probably a foreground structure.

\section{Discussion \label{sec:discussion}}

The relative distances between the molecular cloud complexes and
the relative gradients within them are robust results. The absolute distance 
of the 50~km\,s$^{-1}$ cloud complex is determined with a $\sim$20\% error.
This behaviour does not change significantly
for different analytic profiles and different offsets $a$ and $z_{0}$ 
(Sect.~\ref{sec:method}). However, due to the nonlinearity of our reconstruction method
(Fig.~\ref{fig:LOS_int}), the absolute distance of the 20~km\,s$^{-1}$ cloud complex
strongly depends on the model profile. The LOS distance of the darkest subcloud of this
complex varies between $-90$~pc using Eq.~(\ref{eq:alt2}) and $a=-150$~$\mu$Jy/arcsec$^{2}$
and $-16$~pc using Eq.~(\ref{eq:alt1}) and $a$=0. Since the profile of Eq.~(\ref{eq:alt2})
overestimates the K$_{\rm S}$ band continuum emission (Fig.~\ref{fig:kprofiles}),
we think that the these distances are too low. The main uncertainty comes from our
ignorance regarding the sky subtraction, i.e. the absolute value of the K$_{\rm S}$
band intensity. We have to assume an absolute distance of cloud A 
(Fig.~\ref{fig:gcsketch}). The most plausible scenarios for us are (i) that 
the limit of integration is  $z_{0}=-100$~pc, i.e. the extent of the Nuclear Bulge 
(Launhardt et al. 2002) and $a$=0 and (ii) that $a$=150~$\mu$Jy and $z_{0}=-50$~pc,
which places cloud A at $z < -50$~pc. Both models lead to a very similar
LOS distribution of the giant molecular cloud complexes (Fig.~\ref{fig:LOSprofiles}). 
However, it is not excluded that the LOS distances with respect to Sgr~A$^{*}$
of the 20~km\,s$^{-1}$ cloud complex might be underestimated by up to a factor of 2.

\begin{figure}
	\psfig{file=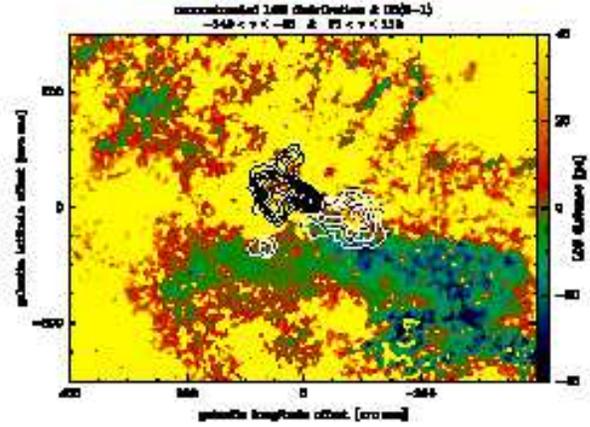,angle=-90,width=\hsize}
	\caption{Contours: IRAM 30m CS(2-1) line observations (G\"{u}sten et al. in prep.). 
	The integrated flux over channels $-140$~km\,s$^{-1} < v < $-31~km\,s$^{-1}$ and
	71~km\,s$^{-1} < v < $115~km\,s$^{-1}$, i.e. the CND, are shown.
	Colors: LOS distance distribution filtered with a median filter of 11 pixels.
	\label{fig:overl_cs2_1_cnd}}
\end{figure}

For a further discussion of the LOS distribution of the GMC complexes in the
Galactic Center we show in Fig.~\ref{fig:overl_cs2_1_cnd} 
IRAM 30m CS(2-1) line observations (G\"{u}sten et al. in prep.). The integrated
flux over channels $-140$~km\,s$^{-1} < v < -31$~km\,s$^{-1}$ and
71~km\,s$^{-1} < v < $115~km\,s$^{-1}$ corresponds to emission from the CND.
The bulk of the CND (except the western arc region)
does not appear in the reconstructed LOS distance distribution.
Since the CND does not cause absorption, its area filling factor must be very low.
With an effective resolution of $\sim 3''$ (0.12~pc) 
this sets an upper limit for the cloud sizes of
$\sim$0.06~pc. This is consistent with the findings of Jackson et al. (1993) who
gave an upper limit of 0.1~pc.

It is a surprise that the Western Arc (Lacy et al. 1991) can be recognized in
our LOS reconstruction (Fig.~\ref{fig:overl_1.2mm}) at the right distance, 
i.e. $d \sim -(1 {\rm -} 2)$~pc (E in Fig.~\ref{fig:gcsketch}).
This means that the clouds that are located in the Western Arc have a larger 
filling factor than those in the rest of the CND. We might speculate that
this is linked to the mechanism that forms an inner edge proposed by
Vollmer \& Duschl (2001). They proposed a scenario where a clump of an external
GMC is falling onto the CND. Clumps that have a low central density become
stretched by the tidal shear. Thus, their area filling factor increases. This is
one possibility to explain the K$_{\rm S}$ band absorption produced by the Western Arc.

\section{Conclusions \label{sec:conclusions}}

We use 2MASS K$_{\rm S}$ images of the Galactic Center region to calculate
the LOS distribution of the GMCs located within the inner 60~pc of the Galaxy.
Using the H band image we distinguish two populations of point sources:
a blue and a red population. The blue population represents a homogeneous 
screen of foreground stars and has to be subtracted from the K$_{\rm S}$ band
image. We reconstructed the line-of-sight distance distribution assuming 
(i) an axis-symmetric stellar distribution and (ii) that the clouds are
optically thick and have an area filling factor $\sim$1, 
i.e. that they block entirely the light from the stars
located behind them. Due to the method of reconstruction, the LOS distances
close to Sgr~A$^{*}$ ($-10\ {\rm pc}< d < 10\ {\rm pc}$) have a small uncertainty, 
whereas it is not excluded that those of larger distances might be underestimated by 
up to a factor of 2. The relative distances are robust results. We conclude that
\begin{itemize}
\item
all structures seen in the 1.2~mm observations (Zylka et al. 1998) and CS(2-1)
observations (G\"{u}sten et al. in prep.) are present in absorption.
\item
the 50~km\,s$^{-1}$ cloud complex is located between 0~pc and $-5$~pc with an 
uncertainty of $\sim$20\%, i.e. in front of
Sgr~A$^{*}$. It has a small LOS distance gradient.
\item
the 20~km\,s$^{-1}$ cloud complex is located in front of the 50~km\,s$^{-1}$ cloud complex.
The subclump of deepest absorption has a LOS distance between $-50$~pc and $-25$~pc.
\item
the 20~km\,s$^{-1}$ cloud complex shows a large LOS distance gradient with galactic longitude.
\item
the Western Arc of the Minispiral has a larger area filling factor than the rest of the CND.
\item
the bulk of the CND is not seen in absorption. This gives an upper limit of the cloud sizes
within the CND of $\sim$0.06~pc.
\end{itemize}

\begin{acknowledgements}
This publication makes use of data products from the Two Micron All Sky Survey, 
which is a joint project of the University of Massachusetts and
the Infrared Processing and Analysis Center/California Institute of Technology, 
funded by the National Aeronautics and Space Administration and the National Science Foundation.
We would like to thank the anonymous referee for helping us to improve
this article significantly.
\end{acknowledgements}

\end{document}